\documentclass[conference]{IEEEtran}

\IEEEoverridecommandlockouts
\usepackage[font=small,labelfont=bf]{caption}
\usepackage{tikz}
\usepackage{eso-pic}
\usepackage{mwe}
\usepackage{graphicx} 
\usepackage{adjustbox} 
\usepackage{booktabs} 
\usepackage{lipsum}  
\usepackage{tikz}
\usepackage{multirow}
\usepackage{acro}
\usepackage{cite}
\usepackage{amsmath,amssymb,amsfonts}
\usepackage{algorithmic} 
\usepackage{textcomp}
\usepackage{xcolor}
\usepackage{hyperref}
\usepackage{cleveref}
\usepackage{siunitx}
\usepackage{graphicx}
\usepackage{float}
\hypersetup{
    colorlinks=true,
    linkcolor=blue,
    filecolor=magenta,      
    urlcolor=magenta,
    citecolor=blue
}
\DeclareAcronym{ai}{short=AI, long= Artificial Intelligence}
\DeclareAcronym{ann}{short=ANN, long= Artificial Neuron Network}
\DeclareAcronym{cnn}{short=CNN, long=Convolutional Neuron Network} 
\DeclareAcronym{dvs}{short=DVS, long=Dynamic Vision Sensor}  
\DeclareAcronym{if}{short=IAF, long= Integrate And Fire} 
\DeclareAcronym{mac}{short=MAC, long= Multiply and Accumulate}  
\DeclareAcronym{nas}{short=NAS, long=Neural Architecture Search}  
\DeclareAcronym{soc}{short=SoC, long=System-on-Chip}   
\DeclareAcronym{eoc}{short=EOG, long= Electrooculography}
\DeclareAcronym{hci}{short=HCI, long=Human–Computer Interactions}

\def\BibTeX{{\rm B\kern-.05em{\sc i\kern-.025em b}\kern-.08em
    T\kern-.1667em\lower.7ex\hbox{E}\kern-.125emX}}
\begin{document}
 
\title{Evaluating Electric Charge Variation Sensors for Camera-free Eye Tracking on Smart Glasses \\
\thanks{
This work was funded by the Innosuisse (103.364 IP-ICT), Swiss National Science Foundation (Grant 219943).} 
}
\author{
    \IEEEauthorblockN{Alan Magdaleno, Pietro Bonazzi, Tommaso Polonelli, Michele Magno}
    \IEEEauthorblockA{ETH Zürich, Zürich, Switzerland} 
}

\maketitle

\AddToShipoutPictureFG*{%
  \AtPageUpperLeft{%
    \begin{tikzpicture}[remember picture,overlay]
      \node[anchor=north] at ([yshift=-2mm]current page.north){%
        \parbox{\paperwidth}{\centering \color{gray}\bfseries
          This paper has been accepted for publication at the\\
          IEEE International Conference on Electronics Circuits and Systems 2025 in Marrakesh, Marocco%
        }%
      };
    \end{tikzpicture}
  }
}

\begin{abstract}
Contactless \ac{eoc} using electric charge variation (QVar) sensing has recently emerged as a promising eye-tracking technique on wearable devices. QVar enables low-power and unobtrusive interaction without requiring skin-contact electrodes. Previous work demonstrated that such systems can accurately classify eye movements using onboard TinyML under controlled laboratory conditions. However, the performance and robustness of contactless \ac{eoc} in real-world scenarios—where environmental noise and user variability are significant—remain largely unexplored. In this paper, we present a field evaluation of a previously proposed QVar-based eye-tracking system, assessing its limitations in everyday usage contexts across 29 users and 100 recordings in everyday scenarios, such as working in front of a laptop. Our results show that classification accuracy varies between 57\% and 89\% depending on the user, with an average of 74.5\%, degrading significantly in the presence of nearby electronic noise sources. Our experimental results show that contactless \ac{eoc} remains viable under realistic conditions, though subject variability and environmental factors can significantly affect classification accuracy. The findings inform the future development of wearable gaze interfaces for human-computer interaction and augmented reality, supporting the transition of this technology from prototype to practice.
\end{abstract} 


\section{Introduction}

Wearable smart glasses are emerging as a next-generation computing platform, integrating sensors and intelligent features to support augmented reality (AR) and other applications~\cite{yin_internet_2023}. A critical functionality for these devices is eye tracking ~\cite{adhanom2023eye}, which can enable intuitive \ac{hci}, context-aware AR interfaces, health monitoring, and hands-free control in assistive technologies~\cite{CHEN2023883, wolf2023eye, yang_aim_2024}. By decoding eye movements~\cite{tinytracker2023}, gaze points ~\cite{tinytracker2023}, and blinks, eye-tracking systems \cite{yang2023wearable} provide valuable insights into user intent and cognitive state. However, implementing eye tracking on everyday wearable glasses poses significant challenges, as traditional techniques often fail to meet the power and comfort constraints of real-world use~\cite{palermo2025advancements}.

An established approach is \ac{eoc} with skin electrodes, it offers lower power consumption and has been applied in \ac{hci} and medical diagnostics. Traditional EOG relies on contact electrodes around the eyes (often wet Ag/AgCl sensors), which can cause discomfort, skin irritation, and require frequent re-calibration due to electrode drift. The need to precisely place sticky electrodes makes standard \ac{eoc} setups unsuitable for everyday smart glasses~\cite{palermo2025advancements}. Therefore, there is strong motivation to develop a non-invasive, user-friendly, and low-power eye-tracking solution that overcomes these shortcomings~\cite{palermo2025advancements}. One promising direction is to leverage the eye’s bioelectric field through contactless \ac{eoc} sensing~\cite{lopez_comparison_2023}. Recent works have explored charge-variation (QVar) sensing~\cite{scharer2024electrasight, crafa2025low}, a technique to measure the subtle quasi-electrostatic field changes produced by eye movements. Unlike classical \ac{eoc}, QVar-based sensors do not strictly require direct skin contact, enabling comfortable and unobtrusive signal acquisition \cite{crafa2025low}. By detecting the corneo-retinal potential (the dipole between the cornea and retina) from a short distance, a QVar sensor on the eyeglass frame can capture horizontal and vertical eye motion signals in a fully non-contact manner (or hybrid, with a mix of contact and contactless electrodes). This approach inherits the low-power advantage of \ac{eoc} while significantly reducing invasiveness. For example, a recent system~\cite{scharer2024electrasight} integrated electrostatic sensors into glasses and demonstrated the classification of nine eye movement types with about 81\% accuracy, validating the feasibility of non-contact eye tracking for multiple gestures. However, prior studies have typically evaluated such eye trackers only in controlled laboratory conditions~\cite{scharer2024electrasight, crafa2025low}. Cross-user variability remains another hurdle – differences in anatomy and signal strength between individuals can degrade accuracy if the system is not tuned per user. In fact, many \ac{eoc}-based interfaces require an elaborate calibration phase for each new user to account for inter-subject differences, which is cumbersome and limits real-world adoption. To date, no solution meets all the requirements of accuracy, low power, comfort, privacy, and user-agnostic operation for eye tracking on everyday glasses \cite{scharer2024electrasight, crafa2025low}.

To address these gaps, this paper presents an eye-tracking field assessment for wearable glasses that uses a QVar-based contactless and contact \ac{eoc} sensing. This approach builds upon our prior work~\cite{scharer2024electrasight, crafa2025low}, in which we introduced a QVar-enabled smart glasses prototype capable of classifying various eye movements with 81\% accuracy using onboard tinyML, without requiring user-specific model tuning~\cite{scharer2024electrasight}. While that earlier system validated the concept of onboard, low-power eye tracking in a controlled setting, it was not evaluated under everyday noisy conditions, and it assumed a one-size-fits-all model. In this work, we extend the evaluation to real-world scenarios and explicitly highlight the issue of inter-subject variability. We also test the lightweight CNN model (54k parameters) deployed on-device for real-time classification in realistic environments – such as during typical laptop use – to examine its robustness during a common application scenario. This work presents one of the first demonstrations of contactless \ac{eoc} eye tracking in an everyday setting, providing concrete evidence of the approach's practical feasibility and limitations.
\section{Related Works} 
QVar is an electrostatic sensing technology well known in the analogical circuit design, but that has been employed for eye tracking only recently~\cite{scharer2024electrasight,das_eog_2024}. In essence, it detects minimal changes in electric charge or quasi-static electric fields in the environment. QVar uses a pair of electrodes to sense variations in the surrounding electric potential. When objects (such as the human body or other materials) move or come into contact, they often generate static charges (triboelectric effect). QVar picks up these charge fluctuations as voltage changes on its electrodes. The sensor’s analog front end is extremely high-impedance, so it can detect even microscopic differential potentials induced between a sensing electrode and a reference. In operation, QVar does not emit any signal of its own, therefore offering very low power consumption, in the range of \qty{100}{\micro\watt}. This passive, capacitance-like sensing mechanism is highly sensitive and cost-effective compared to many conventional sensor principles.

In recent days, a survey about sensing onboard smart glasses \cite{palermo2025advancements} highlighted the importance of non-invasive eye tracking, listed as one of the critical technologies still to be developed. Fernanda Irrera et al. \cite{net:stm} provide the most comprehensive evaluation of QVar to date. Their study introduces a novel QVar‐based methodology for long‐term biopotential recording that simultaneously captures ECG, EEG, \ac{eoc}, and sEMG signals via a multisensor platform with three QVar channels. Each QVar front end draws merely \qty{27}{\micro\watt} during continuous monitoring, and the resulting biopotential waveforms deviate by less than 5 \% from those obtained with clinical‐grade electrodes, demonstrating QVar’s efficacy as a low‐power, ambulatory biopotential acquisition technology. Moreover, in \cite{shi_eye_2023}, authors present a high precision (\ang{5}) eye-tracking system based on electrostatic measurements, fully integrated into glass lenses. Electrostatic measurements based on QVar sensing have not been used only for \ac{eoc}, but several works in the literature address wearable biosensing \cite{manoni_long-term_2022}. However, to the best of our knowledge, these solutions are mainly assessed in a research context based on prototypes tested in controlled scenarios. Therefore, there is a lack of field verification conducted on a multi-subject setup.
 
\definecolor{DeepPurple}{HTML}{981C97}
\definecolor{Green}{HTML}{78BE21}
\definecolor{Red}{HTML}{BA0D2F}
\definecolor{Blue}{HTML}{00629C}
\definecolor{Orange}{HTML}{FFA300}

\section{Methodology} 
\subsection{Hardware Setup}
This work utilizes the ST1VAFE3BX from STMicroelectronics. It includes one charge variation sensing channel and supports in-sensor processing capabilities such as signal filtering, step detection, and finite state machine execution. For the scope of this paper, we employed the custom module designed in \cite{scharer2024electrasight}, providing six QVar channels to enable flexible testing of various electrode placements. Its small footprint (\qty{19.5}{\milli\meter} × \qty{16.5}{\milli\meter}) aligns with the requirements of wearable designs. 

Through extensive experimentation and field trials, two electrode types were selected: one for contact sensing and one for contactless operation. For contact sensing, Softpulse electrodes from Dätwyler—made from conductive elastomer rubber coated with Ag/AgCl—were chosen for their skin compatibility, flexibility, and low impedance, ensuring high-quality signal acquisition. For contactless sensing, ENIG-coated copper sheets were used due to their reliability, ease of integration, and ability to couple well with skin through capacitive effects.

\begin{figure}[t]
    \centering
    \begin{minipage}{0.55\linewidth}
        \centering
        \includegraphics[width=\linewidth]{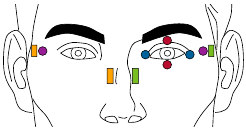}
        \caption*{\textbf{(a)} Electrode placement for the five differential channels in the glasses setup: non-contact: \textcolor{Blue}{horizontal}, \textcolor{Red}{vertical}, \textcolor{DeepPurple}{horizontal corners}; contact: \textcolor{Green}{diagonal left}, \textcolor{Orange}{diagonal right}}
    \end{minipage}%
    \hfill
    \begin{minipage}{0.4\linewidth}
        \centering
        \scriptsize
        \captionof{table}{\textbf{(b)} Class distribution of the full dataset}
        \label{tab:dataset_distribution}
        \begin{tabular}{|l|r|}
            \hline
            \textbf{Class}    & \textbf{Count} \\ \hline
            Left             & 614            \\ \hline
            Right            & 595            \\ \hline
            Up               & 729            \\ \hline
            Down             & 668            \\ \hline
            Down-Left        & 345            \\ \hline
            Up-Left          & 363            \\ \hline
            Up-Right         & 354            \\ \hline
            Down-Right       & 339            \\ \hline
            Blink            & 560            \\ \hline
            Center           & 300            \\ \hline
        \end{tabular}
        \normalsize
    \end{minipage}
    \caption{(a) Electrode placement on the glasses and (b) corresponding class distribution from the dataset collected across 100 analyzed recordings.}
    \label{fig:electrodes_and_distribution}
\end{figure}

The final setup employs five sensing channels, requiring ten electrodes, as depicted in \Cref{fig:electrodes_and_distribution}. Two channels use contact electrodes positioned at the eyeglass nose pads and front temples, forming nearly orthonormal diagonal paths (left and right). The remaining three channels are contactless and use \qty{2.5}{\milli\meter} × \qty{0.7}{\milli\meter} copper sheets placed around the left eye—above and below for vertical detection, and laterally for horizontal tracking—without obstructing the user's vision. Additional electrodes on both temples provide a center channel. Electrode placement was optimized empirically to balance performance with minimal invasiveness.

\subsection{Dataset}
To build a comprehensive and varied dataset, recordings were conducted with 29 subjects, each undergoing 5 separate sessions. During these sessions, subjects were instructed to track a circle displayed on a screen without moving their head or blink. For classification purposes, each instance where the circle changed position was recorded with a timestamp and labeled accordingly. The circle navigated through a total of 8 distinct positions (Up, Down, Left, Right, Down-Left, Down-Right, Up-Left, Up-Right). Additionally, to capture blink events, the word "Blink" would appear in the center of the screen. In order to classify when the eyes are not moving, a "Center" label was added. In total, we collected 100 recordings. 

\subsubsection{Noise Analysis}
A total of 71 recordings were made by the same person in different rooms while seated. Moreover, five other participants contributed ten additional recordings for comparison purposes. Significant differences in measurements were noticed during dataset acquisition, even when recorded with the same subjects. If an electromagnetic wave emitting device was nearby (Laptop, Monitor, Fridge, Lamp, ...), the QVar channels picked up noise. Each session was manually labeled as “clean” or “noisy”. A laptop was nearby for all acquisitions, receiving the raw data from the glasses. In an effort to isolate the influence of the laptop, a test was designed with a single participant wore the QVar-equipped glasses \cite{scharer2024electrasight} and remained stationary (no ocular movements) at several predetermined distances from an idle laptop running only a Python script, and disconnected from its charger. For each distance, the median peak-to-peak value was calculated to quantify baseline noise. 

\subsection{Neural Network Architecture}
Optimal performance was attained with a highly compact CNN comprising only 54'270 trainable parameters. The network ingests a 120-sample window (\qty{0.5}{\second}) from each of the five preprocessed QVar channels. Four consecutive convolutional blocks carry out feature extraction, each block consisting of a one-dimensional convolutional layer with 40 filters, kernel size 5, stride 1, and ReLU activation followed by a subsequent max-pooling layer. The output of the fourth block is flattened and passed through two fully connected layers before a softmax classifier produces probabilities over the ten target classes. 
\section{Experimental Results} 

\subsection{Model performance}
\begin{figure}[b]
    \centering
    \includegraphics[width=1\linewidth]{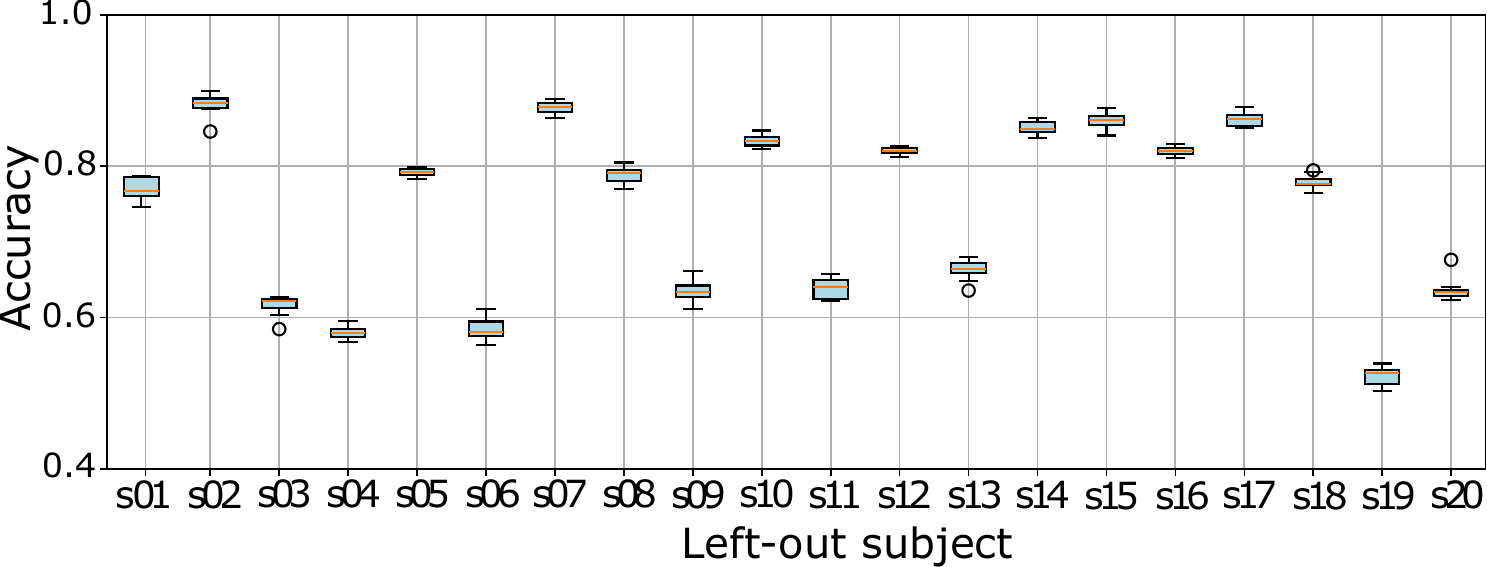}
    \caption{Accuracy on leave-one-subject-out  cross-validation setup. \textit{sXX} indicates the specific subject in the dataset left out during the training and only used for evaluation. This plot highlights the large intra-subject variability on classification accuracy.}
    \label{fig:loso}
\end{figure}

Using the CNN model, an average accuracy of 74.5 \% was achieved when training on 19 out of 20 subjects and testing with data from the left-out subject. \Cref{fig:loso} shows the different test accuracies for every subject. The test accuracy ranged from 57\% to 89\%, indicating substantial inter-subject variability, reaching up to 20\%.

\subsection{Intra-subject Measurement Variability}
\begin{figure*}[t]
    \centering
    \includegraphics[width=1\linewidth]{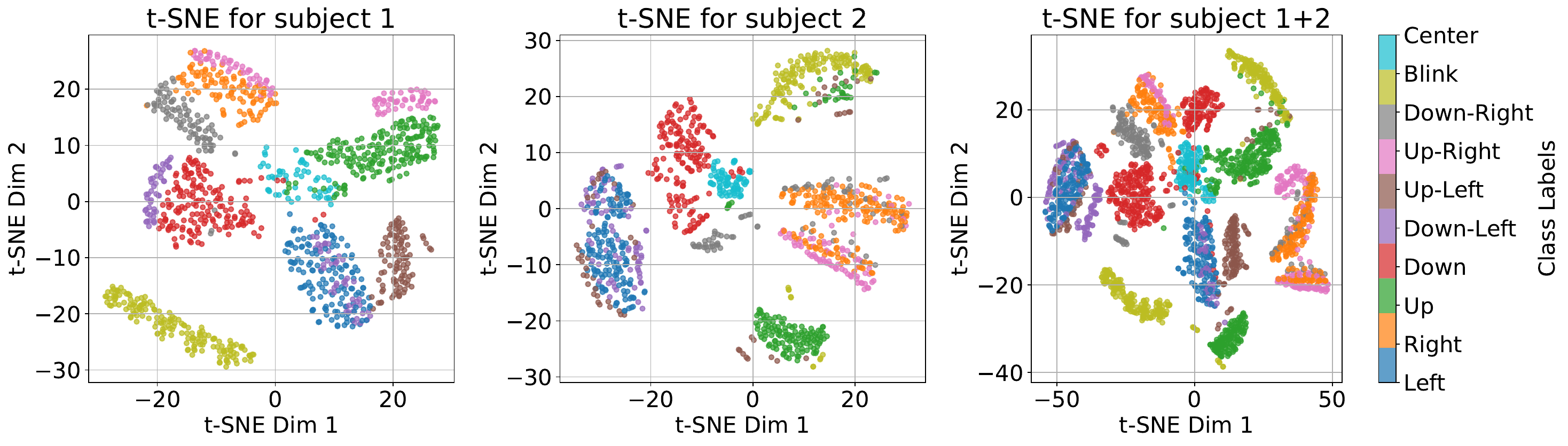}
    \caption{T-SNE projections of Qvar data: (left) Subject 1 only, (middle) Subject 2 only, (right) Subjects 1 \& 2 combined.}
    \label{fig:tsne-comp}
\end{figure*}

The class distributions in feature space were visualized using t-SNE, as shown in  \Cref{fig:tsne-comp}. When visualizing individual subjects (left and middle plots), the embeddings form well-separated clusters for different eye movement classes, indicating that the CNN learns discriminative representations when intra-subject variability is low. However, when data from both subjects are combined (right plot), the inter-class separability degrades, and several class clusters overlap. This merging suggests that inter-subject differences dominate over inter-class differences, which may be due to latent factors such as eye physiology, head movement artifacts, electrode-skin distance, or recording noise patterns that differ between individuals.

\subsection{Signal Quality and Noise Influence}

\begin{figure}[b]
    \centering
    \includegraphics[width=1\linewidth]{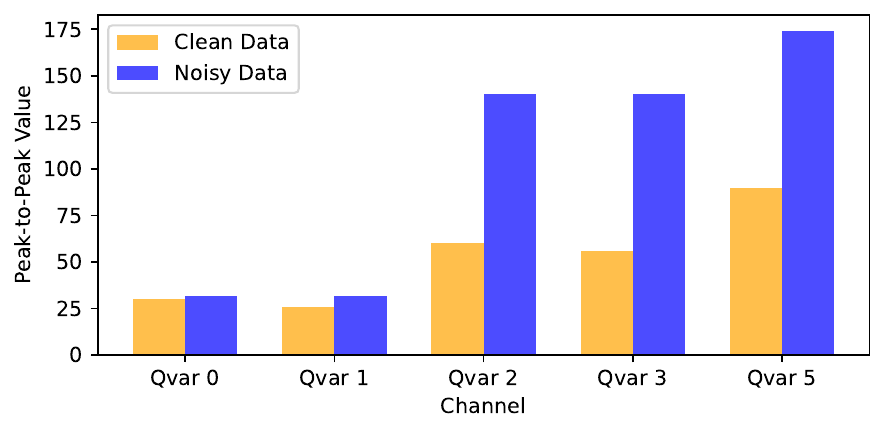}
    \caption{Median peak-to-peak value for each channel. Setup: non-contact: \textcolor{Blue}{horizontal Qvar 0}, \textcolor{Red}{vertical Qvar 1}, \textcolor{DeepPurple}{horizontal corners Qvar 5}; contact: \textcolor{Green}{diagonal left Qvar 2}, \textcolor{Orange}{diagonal right Qvar 3}.}
    \label{fig:peak-peak}
\end{figure}

\begin{figure}[t]
    \centering
    \includegraphics[width=1\columnwidth]{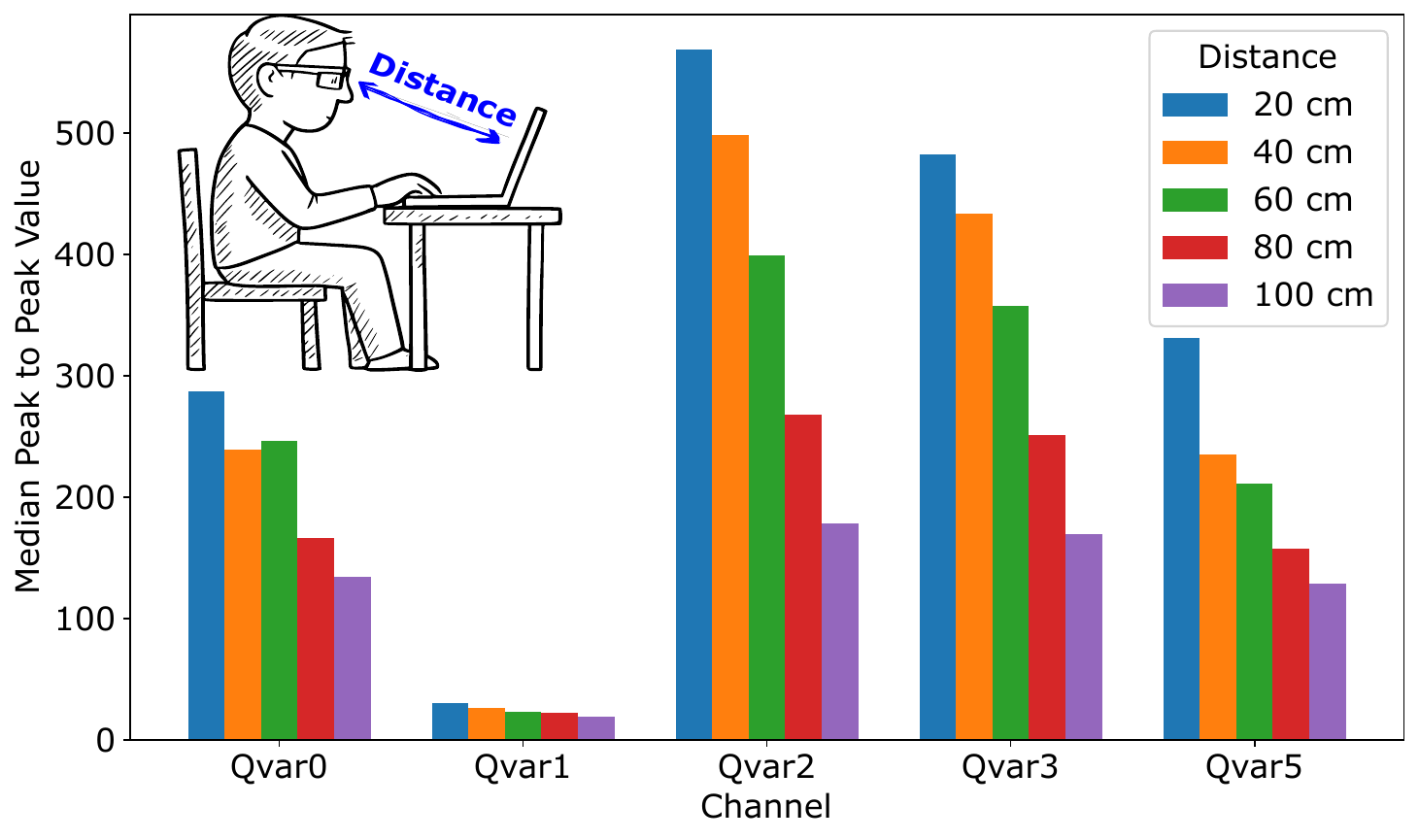}
    \caption{Influence of a laptop on peak-to-peak noise picked-up by the 5 channels. Setup: non-contact: \textcolor{Blue}{horizontal Qvar 0}, \textcolor{Red}{vertical Qvar 1}, \textcolor{DeepPurple}{horizontal corners Qvar 5}; contact: \textcolor{Green}{diagonal left Qvar 2}, \textcolor{Orange}{diagonal right Qvar 3}.}
    \label{fig:laptop}
\end{figure}

\Cref{fig:peak-peak} compares median peak-to-peak values for clean vs. noisy data. Noisy signals, especially in Qvar channels 2 to 5, exhibit significantly higher amplitudes, indicating greater signal distortion. This increased variability can adversely affect classification performance.
Furthermore, as shown in \Cref{fig:laptop}, every channel exhibits an approximately inverse linear relationship between distance and measured noise amplitude. By moving from \qty{20}{\cm} to \qty{100}{\cm} distance, the peak-to-peak value is approximately halved for every channel. 
The higher idle peak-to-peak value indicates a lower overall Signal-to-Noise ratio. The closer a person is to a laptop, the worse the data quality. This behavior confirms that electromagnetic emissions from consumer laptops can measurably degrade Qvar signal quality when operated in close proximity, and underscores the need to control for local electronic devices during data acquisition. 

\section{Conclusion} 
This study provides an in-depth field evaluation of contactless QVar-based \ac{eoc} systems under real-world usage conditions. While prior work demonstrated promising results in controlled environments \cite{scharer2024electrasight,crafa2025low}, our experiments reveal the practical challenges of deploying this technology in everyday settings. Specifically, we observed considerable inter-subject variability in classification accuracy, ranging from 57\% to 89\%, highlighting the difficulty of developing generalizable models for eye movement recognition. The t-SNE analysis further confirmed that signal characteristics are strongly subject-dependent, with diminished class separability when combining data across users. Additionally, we quantified the impact of ambient electromagnetic noise—most notably from consumer electronics like laptops—which significantly reduces signal quality and, consequently, model performance. 
Despite these challenges, our results affirm the viability of contactless \ac{eoc} in realistic conditions and provide crucial insights into the limitations and considerations necessary for robust deployment. These findings pave the way for more adaptive and noise-resilient algorithms, contributing to the broader integration of contactless eye-tracking in wearable human-computer interaction systems.

\bibliographystyle{plain}  
\bibliography{bibliography}

\end{document}